\documentclass[aps,prd,preprint,superscriptaddress,tightenlines,nofootinbib]{revtex4}

\usepackage{graphicx}
\usepackage{dcolumn}
\usepackage{bm}

\begin{document}

\preprint{CLNS 06/1986}       
\preprint{CLEO 06-26}         
\title{A Precision Determination of the $D^0$ Mass}

\author{C.~Cawlfield}
\author{B.~I.~Eisenstein}
\author{I.~Karliner}
\author{D.~Kim}
\author{N.~Lowrey}
\author{P.~Naik}
\author{M.~Selen}
\author{E.~J.~White}
\author{J.~Wiss}
\affiliation{University of Illinois, Urbana-Champaign, Illinois 61801}
\author{R.~E.~Mitchell}
\author{M.~R.~Shepherd}
\affiliation{Indiana University, Bloomington, Indiana 47405 }
% ...
\collaboration{CLEO Collaboration} 
\noaffiliation

\date{January 30, 2007}

\begin{abstract} 

A precision measurement of the $D^0$ meson mass
has been made using  $\sim\!281$ pb$^{-1}$ of $e^{+}e^{-}$ annihilation data 
taken with the CLEO-c detector at the $\psi(3770)$ resonance.
The exclusive decay $D^0\to K_S \phi$ has been used to obtain
$M(D^0)=1864.847\pm0.150(\mathrm{stat})\pm0.095(\mathrm{syst})\;\mathrm{MeV}$.
This corresponds to $M(D^0\overline{D^{*0}})=3871.81\pm0.36$ MeV,
and leads to a well--constrained determination of the binding energy of 
the proposed $D^0\overline{D^{*0}}$ molecule X(3872), as $E_b=0.6\pm0.6~\mathrm{MeV}$.
\end{abstract}
\maketitle

The $D^0\;(c\bar{u})$ and $D^\pm\;(c\bar{d},\;\bar{c}d)$ mesons form 
the ground states of the open charm system.
The knowledge of their masses is important for its own sake,
but a precision determination of the $D^0$ mass has become more important 
because of the recent discovery of a narrow state known as 
X(3872) \cite{belleX,cdfX,d0X,babarX}. 
Many different theoretical models have been proposed  
\cite{x-cc,x-hybrid,x-gb,x-mol} to explain the nature of this state, 
whose present average of measured masses 
is $M({\rm X})=3871.2\pm0.5$ MeV \cite{pdg}.
A provocative and challenging theoretical 
suggestion is that X(3872) is a loosely bound molecule of $D^0$
and $\overline{D^{*0}}$ mesons \cite{x-mol}. This suggestion arises mainly from
the closeness of $M({\rm X}(3872))$ to
$M(D^0)+M(D^{*0})=2M(D^0) + \Delta [M(D^{*0})-M(D^0)]=2(1864.1\pm1.0)+(142.12\pm0.07)~\mathrm{MeV}=3870.32\pm2.0~\mathrm{MeV}$ based on the PDG \cite{pdg} \textit{average} value of the measured
$D^0$ mass, $M(D^0)=1864.1\pm1.0$ MeV. 
This gives the binding energy of the proposed molecule, 
$E_{b}(\mathrm{X}(3872))\equiv M(D^0) + M(D^{*0}) - M({\rm X}(3872)) =-0.9\pm2.1$ MeV.
Although the negative value 
of the binding energy would indicate that X(3872) is not a bound state
of $D^0$ and $\overline{D^{*0}}$, its $\pm$2.1 MeV error does not preclude this possibility.
It is necessary to measure the masses of both $D^0$ and X(3872) with much
improved precision to reach a firm conclusion.
In this Letter we report on a precision measurement of the $D^0$ mass, and
provide a more constrained value of the binding energy of X(3872) 
as a molecule.

Several earlier measurements of the $D^0$ mass exist \cite{pdg}.
The only previous measurements in which sub-MeV precision was claimed 
are the SLAC measurements of
$e^{+}e^{-}\to \psi(3770)\to D^0\overline{D^0}$ by the lead glass 
wall (LGW) \cite{mark1} and the Mark II \cite{mark2} collaborations, 
and the CERN measurement by the
NA32 experiment with 230 GeV $\pi^{-}$ incident on a copper target \cite{na32}.
All three measurements determined the $D^0$ mass using $D^0\to K^-\pi^+$ and $D^0\to K^-\pi^+\pi^-\pi^+$ (and charge conjugates) decays.  
In the SLAC measurements the beam constrained mass was determined as 
$M^{2}(D^0)=E^2_{\mathrm{beam}}-p^2_D$.
The results were $M(D^0)$ = 1863.3 $\pm$ 0.9 $\mathrm{MeV}$ (LGW~\cite{mark1}), and $M(D^0)$ = 1863.8 $\pm$ 0.5 $\mathrm{MeV}$ (Mark~II~\cite{mark2}).  The NA32 experiment reported $M(D^0)$ = 1864.6 $\pm$ 0.3($\mathrm{stat}$) $\pm$ 1.0($\mathrm{syst})~\mathrm{MeV}$ from a simultaneous fit of the mass and lifetime of $D^0$ in the two decays, with the main contribution to the systematic uncertainty arising from magnetic field calibration.
The PDG \cite{pdg} lists the resulting average $D^0$ mass based on the 
measured $D^0$ masses as $M(D^0)_{\mathrm{AVG}}$=1864.1 $\pm$ 1.0 MeV.  
They also list a fitted mass, 
 $M(D^0)_{\mathrm{FIT}}$=1864.5 $\pm$ 0.4 MeV,
based on the updated results of measurements of $D^\pm$, $D^0$, 
$D^\pm_s$, $D^{*\pm}$, $D^{*0}$, and $D_s^{*\pm}$ masses and mass differences.

We analyze $\sim281$ pb$^{-1}$ of $e^{+}e^{-}$ annihilation data 
taken at the $\psi(3770)$ resonance at the Cornell Electron Storage Ring (CESR) with the CLEO-c detector
to measure the $D^0$ mass using the reaction
\begin{equation}
\psi(3770)\to D^0\overline{D^0},\; D^0 \to  K_S\phi,\;K_S\to\pi^+\pi^-,\;\phi\to K^+K^-.
\end{equation}
Our choice of the $D^0\to K_S\phi$ decay mode is motivated by several 
considerations. Our  determination of the $D^0$ mass 
does not depend on the precision of the determination  of the beam energy.
Since $M(\phi)+M(K_S) = 1517$ MeV is a
substantial fraction of $M(D^0)$, the final state particles have small
momenta and the uncertainty in their measurement makes a small contribution
to the total uncertainty in $M(D^0)$. 
This consideration favors $D^0\to K_S\phi$ decay over the more prolific
decays $D^0\to K\pi$ and $D^0\to K\pi\pi\pi$, in which the decay 
particles have considerably larger momenta and therefore greater 
sensitivity to the measurement uncertainties.
An additional advantage of the $D^0\to K_S\phi$ reaction is that
in fitting for $M(D^0)$ the mass of $K_S$ can be constrained to its value 
which is known with precision \cite{pdg}.

The CLEO-c detector \cite{CLEOcDetector} consists of a CsI(Tl) electromagnetic calorimeter, an inner vertex drift chamber, a central drift chamber, and a ring imaging Cherenkov (RICH) detector inside a superconducting solenoid magnet providing a 1.0 Tesla magnetic field.  For the present measurements, the important components are the drift chambers, which provide a coverage of 93\% of $4\pi$ for the charged particles.  The final state pions and kaons from the decays of $K_S$ and $\phi$ have momenta less than 600 MeV/$c$, and they are efficiently identified using measurements of track vertices and ionization loss ($dE/dx$) in the drift chambers.  The detector response was studied using a GEANT-based Monte Carlo simulation \cite{GEANTMC}.

We select $D^0$ candidates using the standard CLEO D--tagging criteria, 
which impose a very loose requirement on the beam energy constrained 
$D^0$ mass, as described in Ref.~\cite{dtag}.  
We select well-measured tracks by requiring that they be fully contained in the barrel region ($|\cos\theta| < 0.8$) of the detector, have transverse momenta $>120$ MeV/$c$, and have specific ionization energy loss, $dE/dx$, in the
drift chamber consistent with pion or kaon hypothesis within 3 
standard deviations. 
 For the pions from $K_S$ decay, we make the additional requirement that they originate from a  common vertex displaced from the interaction point by more than 10 mm. We require a $K_S$ flight distance significance of more than 3
standard deviations.
 We accept $K_S$ candidates with mass in the range $497.7\pm12.0$ MeV.  In addition, for the $K_S$ candidates from the exclusive reaction $D^0\to K_S\phi$, we perform a mass-constrained (1C) kinematic fit and accept in our final sample $K_S$ with $\chi^2<20$.  The $\pi^+\pi^-$ invariant mass distribution is shown in the upper panel of Fig.~1 with a fit to a sum of two Gaussians. 
The fit results are: $M(K_S)$=497.545 $\pm$ 0.112 MeV, $\chi^2/d.o.f.=0.6$, and
full width at half maximum, FWHM = 5.0 MeV.
While the fit is very good, because of the limited statistics the 
resulting $M(K_S)$ does not have the 
precision required for testing the calibration of the detector.  As described later, we use the large statistics data for the inclusive $K_S$ production, 
$D\to K_S+X$, for that purpose.  The lower panel of Fig.~1 shows the $K^+K^-$ invariant mass distribution.
The data are fitted with a Breit-Wigner of width 
$\Gamma(\phi)$ = 4.26 MeV \cite{pdg} convoluted with the Monte Carlo determined
Gaussian with FWHM = 2.8 MeV, and
a linear background. The fit results in
$M(\phi)$ = 1019.518 $\pm$ 0.243 MeV, $\chi^2/d.o.f.=1.1$.
We select events containing a $\phi$ by requiring that $M(K^+K^-)$ of the candidate kaons is within $\pm15~\mathrm{MeV}$ of the 
value $M(\phi)=1019.46$ MeV \cite{pdg}.

\begin{figure}[!t]
\includegraphics*[width=3.5in]{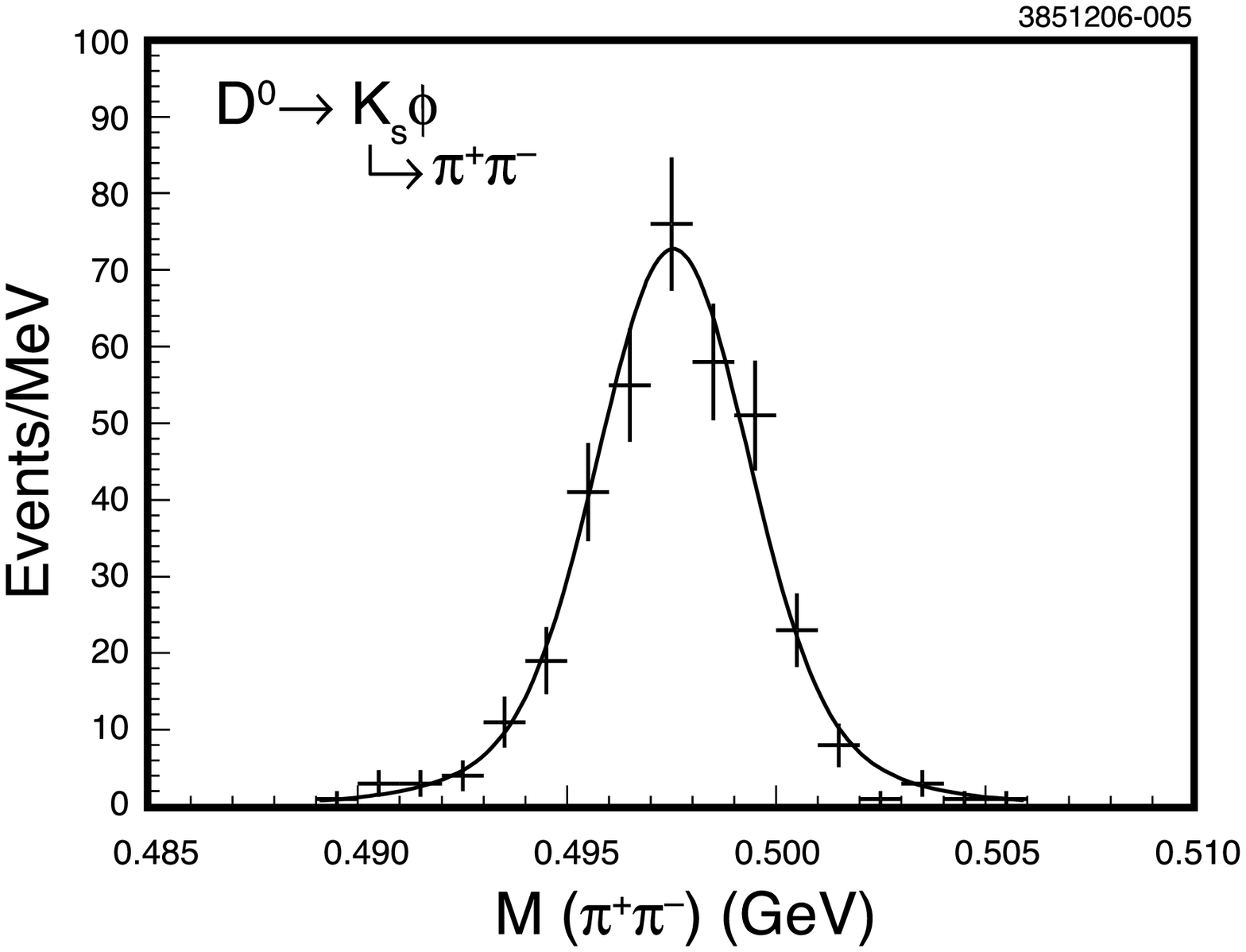}
\includegraphics*[width=3.5in]{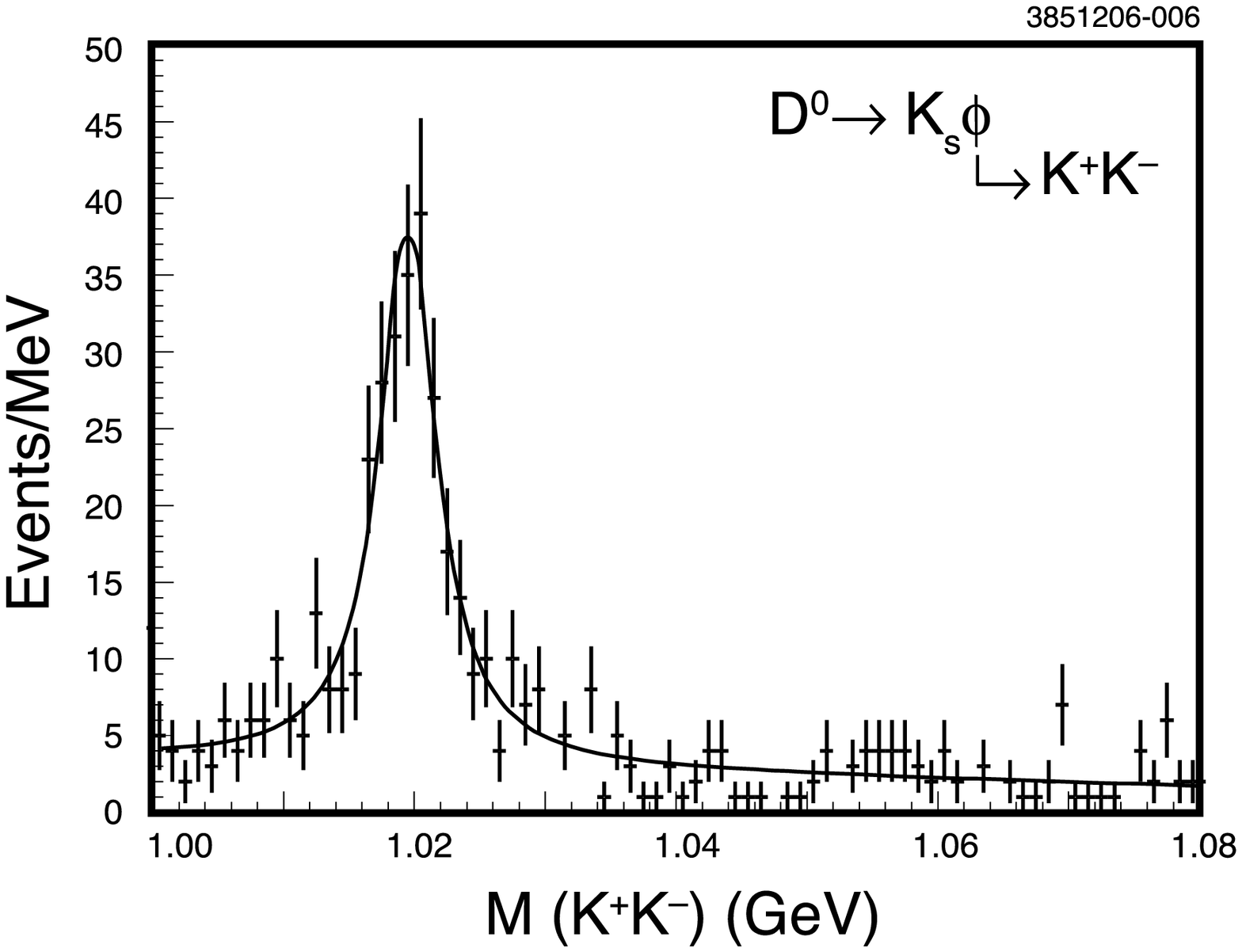}
\caption{Upper plot: Invariant mass of the ($\pi^{+}\pi^{-}$) system for 
$K_S$ decay candidates. The curve shows the fit with the peak shape given 
by the sum of two Gaussians. Lower plot: Invariant mass of the 
($K^{+}K^{-}$) system.
The curve shows the fit with a Breit-Wigner convoluted with a
Gaussian shape and a linear background.}
\end{figure}

Figure 2 shows the invariant mass spectrum of the $D^0$ candidates constructed with $K_S$ and $\phi$ as identified above.  A likelihood fit of the
data in the region 1840--1890 MeV was done with a Gaussian peak and a 
constant background. 
An excellent fit is obtained with the number of fitted events
$N(D^0)=319\pm18$, $\sigma=2.52\pm0.12$ MeV (FWHM = 5.9 MeV), 
$\chi^2/d.o.f.=0.7$, and 
\begin{equation}
M(D^0)=1864.847\pm0.150(\mathrm{stat})~\mathrm{MeV}.
\end{equation}

\begin{figure}[!t]
\includegraphics*[width=3.5in]{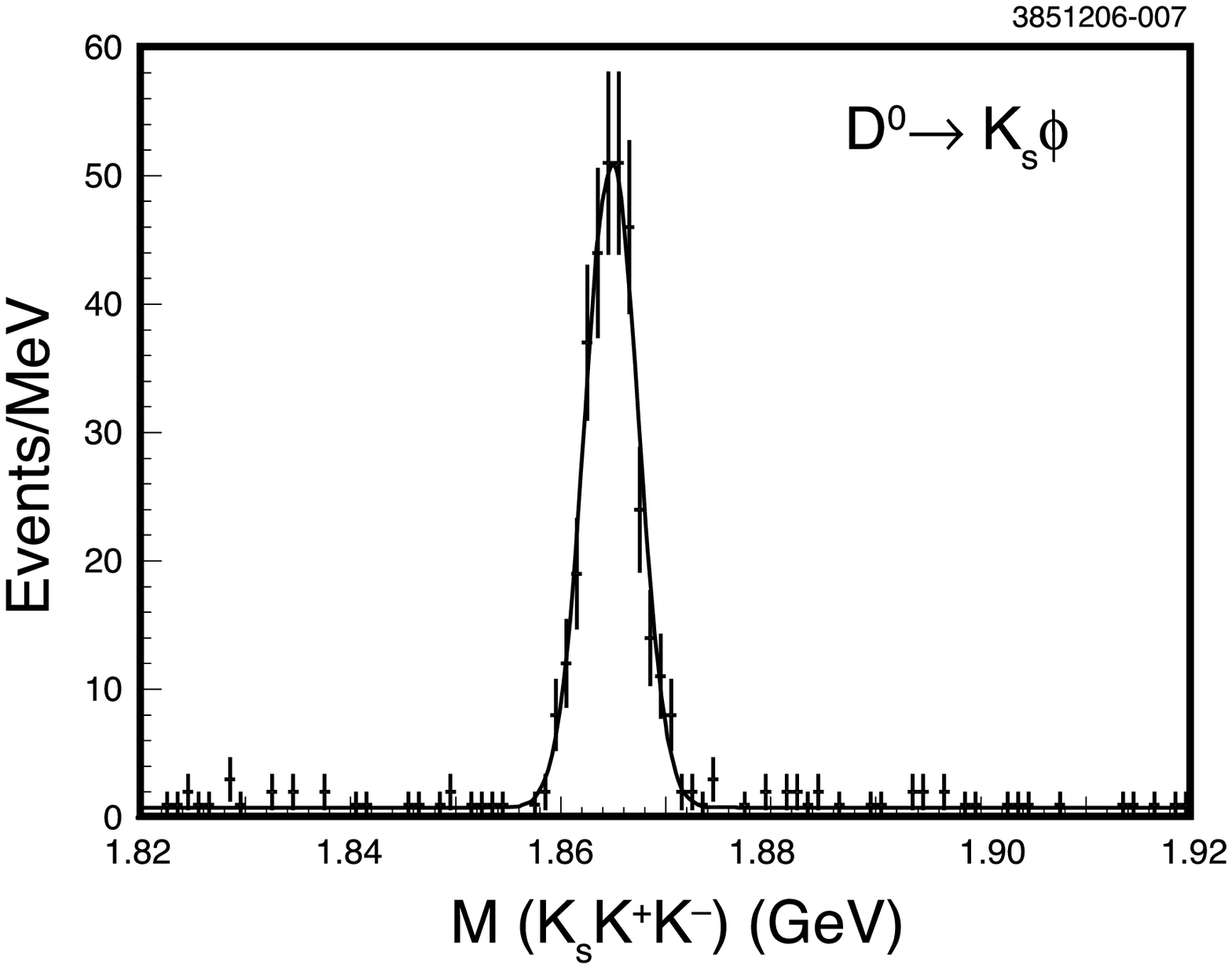}
\caption{Invariant mass of $K_S K^{+}K^{-}$ system for
$D^{0} \to K_S\phi$ decay candidates.
The curve shows fit results with a Gaussian peak shape and a constant 
background.}
\end{figure}

The key to the precision measurement of the $D^0$ mass is in determining 
the accuracy in the detector calibration which can be studied by 
constructing  $M(K_S)$ and $M(\phi)$ from the measured momenta 
of the final state particles, $\pi^\pm$ and $K^\pm$.
We find that $M(\phi)$ is not very sensitive to these variations, because the $K^\pm$ have very small momenta in the rest frame of the $\phi$.  On the other hand, $M(K_S)$ is quite sensitive to the uncertainty in the relatively larger momenta of $\pi^\pm$ in the rest frame of the $K_S$.  The sensitivity of $M(D^0)$ is also large as a consequence of the sensitivity of $M(K_S)$.
We therefore conclude that $M(K_S)$ can be best used to determine 
the accuracy of the detector calibration. As mentioned before, the exclusive 
sample of $D^0\to K_S\phi$ events does not yield a statistically useful result 
for $M(K_S)$. It is possible to determine $M(K_S)$ with much higher 
statistical precision using inclusive $K_S$ production in $D$ decays, 
$D\to K_S+X$.  
Inclusive $K_S$'s were selected from each event that had at least one
candidate $D$ decay. 
The $K_S$ mesons from the decays $D^0\to K_S\phi$ have momenta in the range of
$p(K_S)\approx 0.40-0.65$ GeV/$c$.  
We therefore determine $M(K_S)$ for this range of $p(K_S)$ in the 
inclusive decays.

\begin{figure}[!t]
\includegraphics*[width=3.5in]{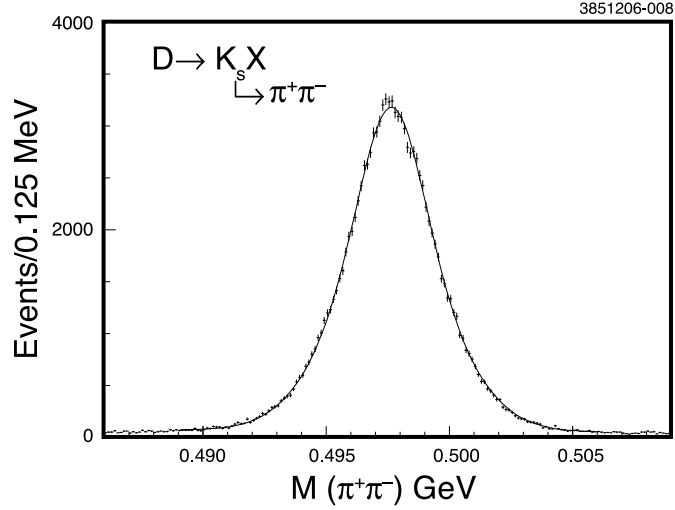}
\caption{Invariant mass of ($\pi^{+}\pi^{-}$) system for 
$K_S$ decay candidates from inclusive sample.
The curve shows fit results with the peak shape given by the sum of two  
Gaussians, and a linear background.}
\end{figure}

Figure 3 shows the $M(\pi^+\pi^-)$ distribution for the inclusive reaction, 
with $p(K_S)$ in the range $0.40-0.65$ GeV/$c$. A fit with the peak shape 
given by the sum of two 
Gaussians and a linear background returns 
\begin{equation}
M(K_S)=497.648\pm0.007(\mathrm{stat})~\mathrm{MeV}.
\end{equation}
The fit has 115,235 $\pm$ 450 events, $\chi^2/d.o.f.=1.07$, 
and  FWHM = 4.7 MeV. 

In order to estimate the systematic error in the above determination of $M(K_S)$, we have studied the variation of $M(K_S)$ as a function of several 
observables associated with $K_S$: $p$($\pi^{\pm}, K_S$), 
$p_T$($\pi^{\pm}$), $p_L$($\pi^{\pm}$), flight distance($K_S)$, 
flight significance($K_S$), 
$\cos(\theta)$($\pi^{\pm}, K_S)$, and $\pi^+\pi^-$ opening angle.  
The largest variation in $M(K_S)$ was found 
with respect to the variation in $\cos(\theta)$ and $p_T$ of $\pi^+$. 
The observed variations contribute a $\pm28$ keV systematic uncertainty
in our determination of $M(K_S)$.  

It is found that Monte Carlo events have a reconstructed output $M(K_S)$ which 
differs by $\pm$21 keV from the input value of $M(K_S)$.
In addition, we determine systematic uncertainties for
different peak fitting procedures: 
$\pm9~\mathrm{keV}$ from variation of the peak shape, 
$\pm1~\mathrm{keV}$ from variation of bin size from 62~keV to 250~keV, 
and $\pm8~\mathrm{keV}$ from variation of fitting range from 
15~MeV to 20~MeV. 
Thus, added in quadrature, the total systematic uncertainty in $M(K_S)$ from the inclusive data is $\pm37~\mathrm{keV}$, and our final result is
$$M(K_S)=497.648\pm0.007(\mathrm{stat})\pm0.037(\mathrm{syst})~\mathrm{MeV}.$$
Since $M(K_S)_{\mathrm{PDG}}=497.648\pm0.022$ MeV, 
$$ M(K_S) - M(K_S)_{\mathrm{PDG}} = 0.000\pm0.044~\mathrm{MeV}.$$
To be conservative, we consider the above maximum difference 
$\pm44$ keV to be a reflection of the possible 
uncertainty in the momentum calibration of the detector, which likely 
arises from uncertainty in the magnetic field calibration and uniformity.
The $B$--field of the CLEO-c detector is set by scaling a map of the $B$--field
such that the measured mass of $J/\psi\to \mu^{+}\mu^{-}$ lies at the 
mass of $J/\psi$ \cite{pdg}.
We have tried several different ways to impose ad-hoc changes in the measured momenta of the pions to produce a $\pm44~\mathrm{keV}$ change in $M(K_S)$ in the \textit{inclusive} data.  We find that when these same changes are applied to the measured momenta of all $\pi^\pm$ and $K^\pm$ in the \textit{exclusive} data, in all cases the change in $M(D^0)$ is nearly twice as large as the change in $M(K_S)$.  We therefore assign $\pm90~\mathrm{keV}$ as the uncertainty in $M(D^0)$ due to the uncertainty in the momentum calibration of the detector.

An independent confirmation of this conclusion is obtained by measuring the mass of $\psi(2S)$ via the reaction $\psi(2S)\to\pi^+\pi^-J/\psi$, which produces $\pi^\pm$ with nearly the same momenta as $\pi^+\pi^-$ and $K^+K^-$ from the $D^0$ exclusive data.  A sample of CLEO-c data for 
$\psi(2S)\to\pi^+\pi^-J/\psi$, $J/\psi\to \mu^{+}\mu^{-}$ was analyzed with 
the track selection and fitting procedure similar to those used to determine $M(D^0)$.  
A mass-constrained kinematic fit for $J/\psi$ was performed, similar to
that done for the $K_S$ in our $D^{0}$ decay.
The fit resulted in $M(\psi(2S))=3686.122\pm0.021~\mathrm{MeV}$.  This differs from the most precise measurement of $M(\psi(2S))=3686.111\pm0.027~\mathrm{MeV}$ by the KEDR collaboration \cite{kedr} by $\Delta M(\psi(2S))=11\pm34~\mathrm{keV}$. Since the detector $B$--field was calibrated at $J/\psi$,
this difference can be attributed to the uncertainty in measurement of
$\pi^{+}\pi^{-}$ momenta, just as in the case of $\pi^{+}\pi^{-}$ in 
inclusive $K_S$.
This assures us that our assignment of $\pm90~\mathrm{keV}$ as the systematic uncertainty in $M(D^0)$ due to detector calibration is conservative. 

\begin{table}
\begin{center}
\begin{tabular}{lc}
\hline 
 &Systematic Error(MeV) \\
\hline
Detector Calibration  & $\pm0.090$   \\
Monte--Carlo input/output & $\pm0.022$  \\
Bin size (0.002--2 MeV) & $\pm0.018$  \\
Unbinned fit    & $\pm0.007$ \\
Peak Shape(single/double Gaussian) & $\pm0.003$  \\
Background Shape (const./linear) & $\pm0.007$  \\
Fit interval ($\pm$20 MeV) & $\pm0.002$  \\
\hline
Sum in Quadrature &  $\pm0.095$ \\
\hline
\end{tabular}
\end{center}
\caption{Summary of systematic errors in $M(D^0)$.}
\end{table}

Other contributions to systematic errors in $M(D^0)$ are smaller, and are
listed in Table I.

Thus, our final result is
\begin{equation}
M(D^0)=1864.847\pm0.150(\mathrm{stat})\pm0.095(\mathrm{syst})\;\mathrm{MeV}.
\end{equation}
Adding the errors in quadrature, we obtain
\begin{equation}
M(D^0)=1864.847\pm0.178~\mathrm{MeV}.
\end{equation}
This is significantly more precise that the current PDG average \cite{pdg}.

Our result for $M(D^0)$ leads to $M(D^0\overline{D^{*0}})$ = 3871.81 $\pm$ 0.36 MeV.
Thus, the binding energy of X(3872) as a $D^0\overline{D^{*0}}$ 
molecule is $E_b=(3871.81\pm0.36)-(3871.2\pm0.5)=+0.6\pm0.6~~\mathrm{MeV}.$
This result provides a strong constraint for the theoretical predictions for the decays of X(3872) if it is a $D^0\overline{D^{*0}}$ molecule \cite{x-mol}.
The error in the binding energy is now dominated by the error in the X(3872) 
mass measurement, which will hopefully improve as the results from the analysis of larger luminosity data from various experiments become available.

We gratefully acknowledge the effort of the CESR staff
in providing us with excellent luminosity and running conditions.
This work was supported by
the A.P.~Sloan Foundation,
the National Science Foundation,
the U.S. Department of Energy, and
the Natural Sciences and Engineering Research Council of Canada.


\begin{thebibliography}{99}

\bibitem{belleX}  Belle Collaboration, S. K. Choi \textit{et al.},  
        Phys. Rev. Lett. {\bf 91}, 262001 (2003).

\bibitem{cdfX} CDF II Collaboration, D. Acosta \textit{et al.},   
        Phys. Rev. Lett. {\bf 93}, 072001 (2004).

\bibitem{d0X} D{\O} Collaboration, V. M. Abazov  {et al.}, 
        Phys. Rev. Lett. \textbf{93}, 162002 (2004).

\bibitem{babarX}  {\slshape{B{\scriptsize{A}}B{\scriptsize{AR}}}} 
        Collaboration, B. Aubert \textit{et al.},  Phys. Rev. \textbf{D 71}, 071103 (2005).

\bibitem{x-cc} E. J. Eichten, K. Lane, and C. Quigg, Phys. Rev. Lett. {\bf 89}, 162002 (2002); Phys. Rev. {\bf D 69}, 094019 (2004); T. Barnes and S. Godfrey, Phys. Rev. {\bf D 69}, 054008 (2004).

\bibitem{x-hybrid} F. E. Close and P. R. Page, Phys. Lett. \textbf{B 578}, 119 (2004).

\bibitem{x-gb} K. K. Seth, Phys. Lett. \textbf{B 612}, 1 (2005).

\bibitem{x-mol} E. S. Swanson, Phys. Lett. \textbf{B 588}, 189 (2004); N. A. T\"{o}rnqvist, Phys. Lett. \textbf{B 599}, 209 (2004);
M. B. Voloshin, Phys. Lett. \textbf{B 579}, 316 (2004).


\bibitem{pdg} Particle Data Group, W.-M. Yao \textit{et al.}, J. Phys. G: Nucl. Part. Phys. \textbf{33}, 1 (2006).


\bibitem{mark1} I. Peruzzi \textit{et al.},
        Phys. Rev. Lett. \textbf{39}, 1301 (1977).

\bibitem{mark2} R.H. Schindler \textit{et al.},
        Phys. Rev. \textbf{D 24}, 78 (1981).

\bibitem{na32} ACCMOR Collaboration, S. Barlag \textit{et al.},  Zeit. f\"ur Phys. \textbf{C 46},  563 (1990).

\bibitem{CLEOcDetector}
G. Viehhauser,
Nucl. Instrum. Meth. \textbf{A 462}, 146 (2001);
D. Peterson \textit{et al.}, 
Nucl. Instrum. Meth. \textbf{A 478}, 142 (2002). 

\bibitem{GEANTMC}R. Brun \textit{et al.}, 
	CERN Long Writeup W5013 (1994), unpublished.

\bibitem{dtag} CLEO Collaboration, Q. He \textit{et al.}, Phys. Rev. Lett. \textbf{95}, 121801 (2005).

\bibitem{kedr} KEDR Collaboration, V. M. Aulchenko \textit{et al.}, Phys. Lett. \textbf{B 573}, 63 (2003).

\end{thebibliography}
\end{document}